\documentclass[aps,prl,twocolumn,groupedaddress]{revtex4-1}
\usepackage{units}
\usepackage{amsmath}
\usepackage{amssymb,amsfonts}
\usepackage{graphicx}
\usepackage{bm}
\usepackage{multirow,color,relsize,microtype}%ulem
\usepackage{dcolumn}% Align table columns on decimal point
\usepackage{xcolor}

\newcommand{\be}{\begin{equation}}
\newcommand{\ee}{\end{equation}}

\newcommand{\cc}[1]{{\color{black}#1}}
\newcommand{\e}{\omega}
\newcommand{\re}[1]{\text{Re}\left[#1\right]}

\begin{document}

\title{Pseudo-chirality: a manifestation of Noether's theorem in non-Hermitian systems}

\author{Jose D. H. Rivero}
\affiliation{\textls[-18]{Department of Physics and Astronomy, College of Staten Island, CUNY, Staten Island, NY 10314, USA}}
\affiliation{The Graduate Center, CUNY, New York, NY 10016, USA}

\author{Li Ge}
\email{li.ge@csi.cuny.edu}
\affiliation{\textls[-18]{Department of Physics and Astronomy, College of Staten Island, CUNY, Staten Island, NY 10314, USA}}
\affiliation{The Graduate Center, CUNY, New York, NY 10016, USA}

\date{\today}

\begin{abstract}
Noether's theorem relates constants of motion to the symmetries of the system. Here we investigate a manifestation of Noether's theorem in non-Hermitian systems, where the inner product is defined differently from quantum mechanics. In this framework, a generalized symmetry which we term pseudo-chirality emerges naturally as the counterpart of symmetries defined by a commutation relation in quantum mechanics. Using this observation, we reveal previously unidentified constants of motion in non-Hermitian systems with parity-time and chiral symmetries. \cc{We further elaborate the disparate implications of pseudo-chirality induced constant of motion: It signals the pair excitation of a generalized ``particle'' and the corresponding ``hole'' but vanishes universally when the pseudo-chiral operator is anti-symmetric. This disparity, when manifested in a non-Hermitian topological lattice with the Landau gauge, depends on whether the lattice size is even or odd. We further discuss previously unidentified symmetries of this non-Hermitian topological system, and we reveal how its constant of motion due to pseudo-chirality can be used as an indicator of whether a pure chiral edge state is excited.}
\end{abstract}

% insert suggested keywords - APS authors don't need to do this
%\keywords{}

\maketitle

Noether's theorem is a powerful statement that relates the symmetries of the system to its constants of motion or conservation laws. In classical mechanics, both energy and momentum conservation can be derived from Noether's theorem under the invariance of time and spatial translations, using, for example, the Hamiltonian's equation of motion with the Poisson bracket as its special form. Correspondingly, a manifestation of Noether's theorem in quantum mechanics is often expressed using the Ehrenfest's theorem \cite{Shankar} in the Schr\"odinger picture:
\be
\frac{d }{d t}\langle A\rangle = \frac{1}{i\hbar}\langle [A,H] \rangle + \left\langle \frac{d A}{d t} \right\rangle,\label{eq:Ehrenfest}
\ee
where the commutation relation replaces the Poisson bracket. For an operator $A$ without explicit time dependence, it then follows that its expectation value $\langle{A}\rangle$ is a constant of motion when it commutes with the Hamiltonian.

Ehrenfest's theorem is a restatement of the Schr\"odinger equation and the fact that the Hamiltonian in quantum mechanics is Hermitian. Meanwhile, the study of non-Hermitian systems and their unique properties have attracted fast growing interest in the last two decades \cite{NPreview,NPhyreview,NMreview,RMP,EPreview}, especially those empowered by parity-time ($PT$) symmetry \cite{Bender1}. While its ramification in quantum theories is still under intense investigation, its application in different fields has led to a plethora of findings, ranging from nonlinear dynamics \cite{PTnonlinear_book}, atomic physics \cite{antiPT_exp1}, photonics \cite{NPreview}, acoustics \cite{PTacoustic}, microwave \cite{Jordan}, electronics \cite{wireless}, to quantum information science \cite{retrieval}. Subsequent explorations have also revealed other novel non-Hermitian symmetries, including, for example, anti-PT symmetry \cite{antiPT,antiPT_exp4}, odd-time-reversal PT symmetry \cite{Mathur,Konotop}, non-Hermitian particle-hole (NHPH) symmetry \cite{zeromodeLaser,Defect,Malzard,Kawabata}, and non-Hermitian chiral symmetry \cite{zeromodeLaser,NHC_arxiv,Kawabata_prx}. Benefiting from these findings, different devices such as single-mode lasers \cite{Feng,Hodaei}, robust power transfer circuits \cite{wireless}, laser-anti-lasers \cite{Longhi,CPALaser,CPALaser_exp,CPALaser_observation} and on-chip lasers carrying orbital angular momentum \cite{Miao} have been demonstrated, which also shine light on a new type of light-matter interaction \cite{Jordan}.

Being non-Hermitian, the total energy or particle number is not a constant of motion in such systems in general, but the aforementioned symmetries do lead to generalized conservation laws \cite{Znojil,pseudoH1,conservation,conservation2D,ramezani_pra10}. These findings, however, were obtained either using other techniques or by following closely the Hermitian manifestation of Noether's theorem. For example, the generalized flux conservation relation $|1-T|=\sqrt{R_LR_R}$ that relates the reciprocal transmittance $T$ in a one-dimensional $PT$-symmetric waveguide to the directional reflectance $R_{L,R}$ was found by analyzing the structure of the scattering matrix \cite{conservation}. And the conserved pseudo-norms, e.g., the expectation value of either the parity operator \cite{Znojil} or the metric $\eta$ that defines pseudo-Hermiticity \cite{pseudoH1}, were proposed to justify quantum theories with non-Hermitian Hamiltonians, where the inner product between a bra and a ket state is defined using the same convention as in quantum mechanics.

In this Letter, we first propose to study an alternative form of Ehrenfest's theorem, which leads to a previously unknown extension of Noether's theorem in non-Hermitian systems. A generalized symmetry, which we term pseudo-chirality, emerges naturally and replaces the standard symmetry defined by a commutation relation in quantum mechanics. Based on this observation, we identify and analyze previously overlooked constants of motion in non-Hermitian systems. \cc{We start with a simple pseudo-spin Hamiltonian, where the contrast and connection between pseudo-chirality and chiral symmetry are examined. The disparate physical implications of pseudo-chirality induced constant of motion are then elucidated: It signals the pair excitation of a generalized ``particle'' and the corresponding ``hole'' in general, and it vanishes universally when the pseudo-chiral operator is anti-symmetric. We show that this disparity, when manifested in a non-Hermitian topological lattice with the Landau gauge, depends on whether the lattice size is even or odd. We further discuss previously unidentified symmetries of this non-Hermitian topological system,} and we reveal how its constant of motion due to pseudo-chirality can be used as an indicator of whether a pure chiral edge state is excited.

The most natural extension of Ehrenfest's theorem in non-Hermitian systems is to replace a Hermitian $H$ by a non-Hermitian one. The structure of the Ehrenfest's theorem does not change, and one is only required to keep $H^\dagger$ separated from $H$. The result is given by
\be
\frac{d }{d t}\langle A\rangle = \frac{1}{i\hbar}\langle AH-H^\dagger A\rangle + \left\langle \frac{d A}{d t} \right\rangle, \label{eq:Ehrenfest1}
\ee
and we recover Ehrenfest's theorem given by Eq.~(\ref{eq:Ehrenfest}) when $H$ is Hermitian. Here $A$, as all observables are in quantum mechanics \cite{Weinberg}, is a \text{linear} operator that satisfies $A[a\psi+b\psi']=aA\psi+bA\psi'\,(a,b\in\mathbb{C})$. Equation (\ref{eq:Ehrenfest1}) then indicates that for a time-independent operator $A$, its expectation value $\langle A\rangle$ is a constant of motion if $AH = H^\dagger A$, or
\be
H^\dagger = AHA^{-1} \label{eq:pseudoH}
\ee
when $A$ is invertible. This observation led to the introduction of pseudo-Hermiticity in non-Hermitian systems \cite{pseudoH1}, where $A$ is interpreted as a metric.

To explore previously unknown extensions of Noether's theorem in non-Hermitian systems, we first note that there is more than one way to define the inner product in non-Hermitian systems. In our discussion below, we define it without the complex conjugation, i.e.,
\be
( \mu | \nu ) \equiv \psi_\mu^T\psi_\nu.\label{eq:inner2}
\ee
\cc{We focus on systems with an asymmetric Hamiltonian (i.e., $H\neq H^T$), and hence this inner product is \textit{distinct} from the biorthogonal product to be discussed below.} With this inner product, we find that the temporal evolution of  $(A)\equiv(\psi| A|\psi)$ is given by \cite{SM}
\be
\frac{d }{d t}(A) = \frac{1}{i\hbar}( AH+H^TA) + \left( \frac{d A}{d t} \right).\label{eq:Ehrenfest2}
\ee
It then indicates that $(A)$ is a constant of motion when $A$ does not depend on time explicitly and satisfies $AH = -H^T A$, or
\be
H^T = -AHA^{-1} \label{eq:pseudoC}
\ee
when $A$ is invertible.

We refer to the symmetry defined by Eq.~(\ref{eq:pseudoC}) as pseudo-chirality, not just because it involves a similar transformation as in the definition of pseudo-Hermiticity, but also due to the fact that it warrants, as we will show, a symmetric spectrum about the origin of the complex energy plane, similar to the consequence of chiral symmetry in non-Hermitian systems \cite{zeromodeLaser,Malzard,NHC_arxiv}. Here chiral symmetry in both Hermitian \cite{Hasan} and non-Hermitian systems \cite{SM} is defined by $\{H,\Xi\}=0$, and it differs from pseudo-chirality by the absence of the matrix transpose.

This transpose, however, has a profound consequence: unlike the chiral operator $\Xi$ or any standard symmetry operators, $A$ does not transform one (right) eigenstate of $H$ into another. Instead, it maps a right eigenstate of $H$ to one left eigenstate: The left and right eigenstates of a non-Hermitian Hamiltonian are defined by \cite{RMP}:
\be
H\psi_\mu=\e_\mu\psi_\mu,\quad \bar{\psi}_\mu^T H=\e_\mu\bar{\psi}_\mu^T. \label{eq:eig_LR}
\ee
We note that a pair of $\psi_\mu$ and $\bar{\psi}_\mu$ are different in general but share the same eigenvalue $\e_\mu$. In addition, they satisfy the biorthogonal relation
\be
(\bar{\mu}|\nu) \equiv {\bar\psi}_\mu^T\psi_\nu = \delta_{\mu\nu} \label{eq:biorth}
\ee
away from an exceptional point \cite{SM}. The second relation in Eq.~(\ref{eq:eig_LR}) is also often written as $H^T\bar{\psi}_\mu = \e_\mu\bar{\psi}_\mu$, and a Hamiltonian with pseudo-chirality then has the following property:
\be
H^T[A\psi_\mu]=-A H\psi_\mu = -\e_\mu [A\psi_\mu], \label{eq:map}
\ee
which indicates that $A\psi_\mu\equiv c\bar{\psi}_\nu$ is a left eigenstate of $H$ with eigenvalue $\e_\nu=-\e_\mu$, where $c$ is a factor to be determined by normalization \cite{SM}. In other words, the spectrum of $H$ is symmetric about the origin of the complex energy plane, and $A$ maps a right eigenstate of $H$ to one of its left eigenstates. \cc{This unique property also hints at the contrasting behaviors of pseudo-chirality and chiral symmetry in terms of fulfilling Wigner's theorem \cite{Weinberg}, a fundamental proposition regarding possible forms of symmetries in quantum mechanics \cite{SM}.}

\cc{As mentioned before, the biorthogonal relation in Eq.~(\ref{eq:biorth}) is different from the non-Hermitian inner product defined in Eq.~(\ref{eq:inner2}). As a result, they have distinct properties, and most noticeably, $(\bar{\mu}|\mu)$ vanishes at an exceptional point of $\e_\mu$ while $(\mu|\mu)$ does not. These two inner products become the same only when the Hamiltonian is symmetric, and so do the left and right eigenstates. One could define an alternative non-Hermitian expectation value in the same spirit of the biorthogonal product, whose equation of motion however does not lead to a new symmetry \cite{SM}.}

\begin{figure}[t]
\includegraphics[clip,width=\linewidth]{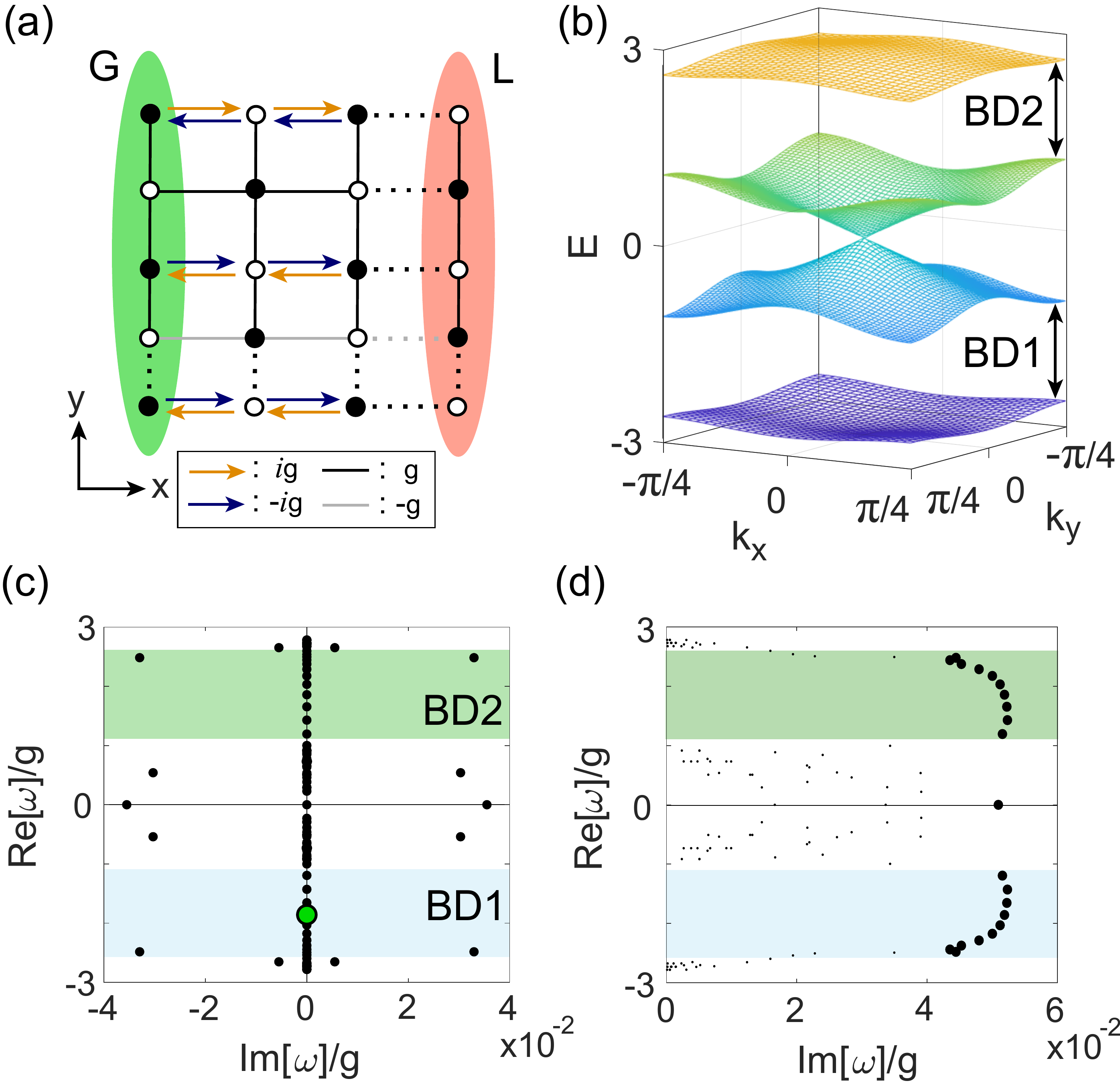}
\caption{Non-Hermitian symmetries in a topological lattice with the Landau gauge. (a) Schematic showing the couplings and the gain (left) and loss (right) edges. (b) Its band structure showing one quarter of the first Brillouin zone. The lattice constant is set to 1. (c) Eigenvalues of $H$ in a 11 by 11 lattice. Clustered dots in BD1 give the CW chiral edge band, and the large dot shows the mode plotted in Fig.~\ref{fig:topo2}(a). (d) Same as (c) but with gain on both edges. High-gain modes are enlarged.
} \label{fig:topo}
\end{figure}

\cc{We start our exploration of constants of motion in non-Hermitian systems by considering the pseudo-spin Hamiltonian $H_s=\vec{b}\cdot\vec{\sigma}$, where $\vec{b}=[b_1,b_2,b_3]$ is an arbitrary complex vector and $\vec{\sigma}=[\sigma_1,\sigma_2,\sigma_3]$ consists of the three Pauli matrices. $H_s$ is Hermitian only when $b_{1,2,3}$ are real, and it gives the most studied form of $PT$-symmetric Hamiltonians when $b_2=0$ and $b_1,ib_3\in \mathbb{R}$ \cite{SM}. The two eigenvalues of $H_s$ are given by
\be
\lambda_1=\sqrt{b_1^2+b_2^2+b_3^2}=-\lambda_2,
\ee
which indicates the presence of either chiral symmetry or pseudo-chirality. In fact, both symmetries are properties of $H_s$, e.g.,
\begin{gather}
\{\Pi,H\}=0,\quad H^T = -AHA^{-1},
\end{gather}
where $\Pi\equiv b_3\sigma_1-b_1\sigma_z$ and  $A=A^{-1}=\sigma_2$.

\begin{figure}[t]
\includegraphics[clip,width=\linewidth]{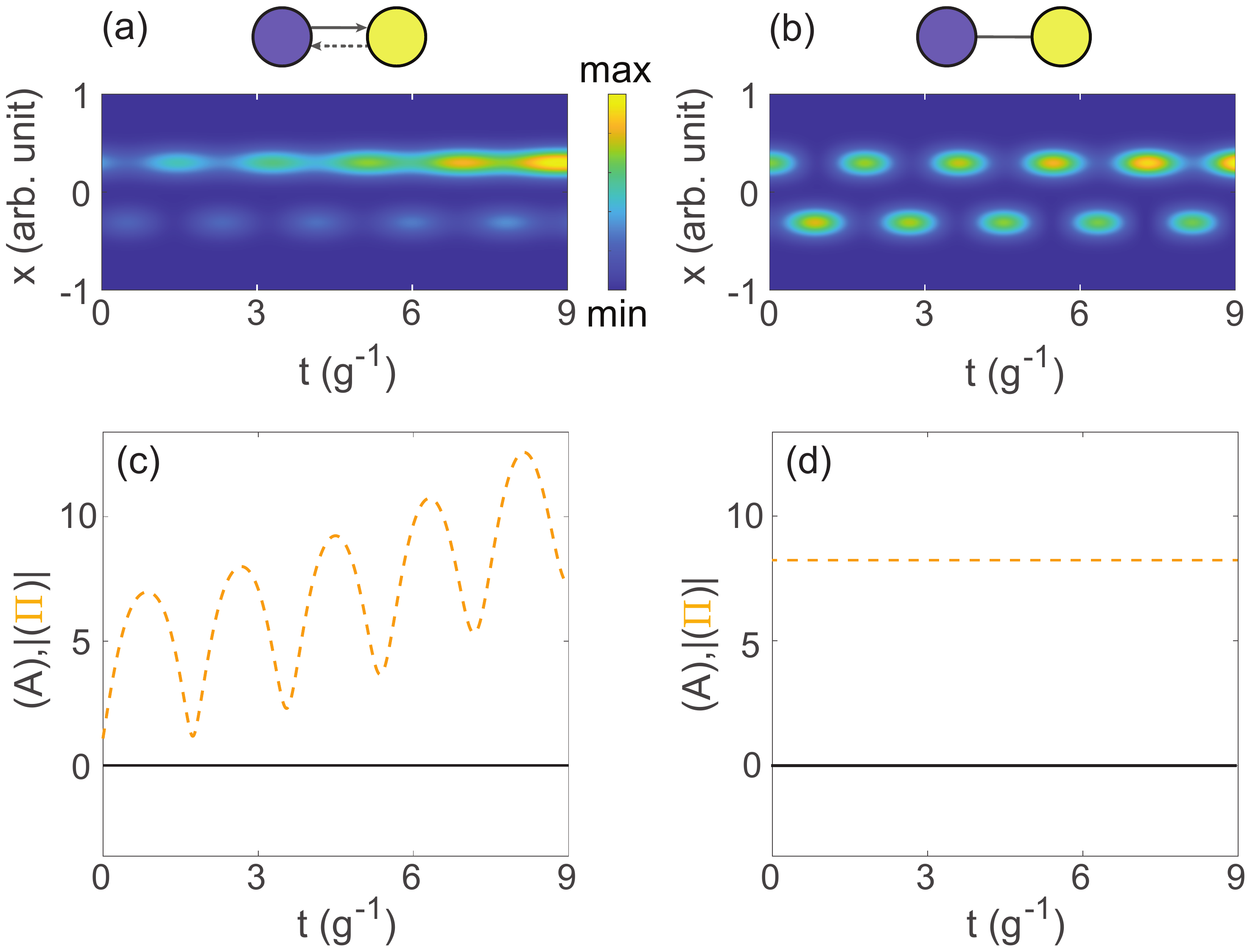}
\caption{Constants of motion due to pseudo-chirality in a non-Hermitian dimer. (a,b) Intensity evolution in the dimer as a function of time. $b_{1,2}=1,1$ in (a) and $-\sqrt{2},0$ in (b), reflected by the asymmetric and symmetric couplings in the insets. In both cases $\psi_1=2,\psi_2=1$ at $t=0$ and $b_3=1+0.1i$. Spatial Gaussian mode profiles are imposed along $x$. (c,d) Constant(s) of motion in the two cases.
} \label{fig:Hs}
\end{figure}

Noether's theorem manifested by Eq.~(\ref{eq:Ehrenfest2}) tells us that $(A)$ is a constant of motion in this system for an arbitrary $\vec{b}$ while $(\Pi)$ is not. We verify this prediction using a set of experiment-friendly parameters in Fig.~\ref{fig:Hs}(a), with real $b_{1,2}$ but leaving $b_3$ complex, representing two complex-detuned oscillators (such as optical waveguides or cavities) and asymmetric couplings with opposite phases. This dimer is one essential building block in non-Hermitian topological systems \cite{topolaser1,topolaser2,toporouting} as we will see later in Fig.~\ref{fig:topo}. If the system (and the coupling) becomes symmetric, i.e., $b_2=0$ [Fig.~\ref{fig:Hs}(b)], $(\Pi)$ is now a conserved quantity as well [Fig.~\ref{fig:Hs}(d)]. This is because $\Pi$ is also a pseudo-chirality operator when the Hamiltonian is symmetric, with the additional transpose of $H$ that differentiates the two symmetries inconsequential now. 

In both cases, we find $(A) = i\psi_2\psi_1 -i \psi_1\psi_2 = 0$ for an arbitrary state $\psi = [\psi_1, \psi_2]^T$. While obvious in this case, it is a \textit{universal} property due to $A^T=-A$. More generally, let us consider a pair of generalized (and charge-neutral) ``particle" $\psi_{\mu}$ and ``hole" $\psi_{\nu}$ with $\e_\mu=-\e_\nu$ that are related by pseudo-chirality through their biorthogonal partners $\bar{\psi}_{\mu,\nu}$. We impose the biorthogonal relation (\ref{eq:biorth}) and write $A\psi_\mu=\bar{\psi}_\nu$, $A\psi_\nu=c\bar{\psi}_\mu$, which leads to
\be
1 = (\bar{\mu}|\mu) = \psi_\nu^T\frac{A^TA^{-1}}{c}\bar{\psi}_\nu. \label{eq:c-1}
\ee
This expression dictates that $c=-1$ if $A^T=-A$, which is seen when compared with $1=(\bar{\nu}|\nu)=\psi_\nu^T\bar{\psi}_\nu$. Now if $\psi$ is an arbitrary superposition of this particle-hole pair, i.e., $\psi=b_\mu\psi_\mu+b_\nu\psi_\nu\, (b_{\mu,\nu}\in\mathbb{C})$, then $(A)=b_\mu b_\nu[(\nu|A|\mu)+(\mu|A|\nu)]=b_\mu b_\nu[(\bar{\nu}|\nu)-(\bar{\mu}|\mu)]=0$. It can be easily checked that this result holds when $\psi$ consists of multiple particle-hole pairs, even when they are degenerate zero modes (i.e., $\e_\mu=-\e_\nu=0$) because Eq.~(\ref{eq:c-1}) does not rely on the energy values.

A disparate behavior is observed for a generic pseudo-chirality operator $A$ (i.e., not anti-symmetric), where $c$ in Eq.~(\ref{eq:c-1}) does not take a specific value. $(A)$ now signals the pair excitation of a generalized particle and the corresponding hole in non-Hermitian systems. From our discussion above, it follows that this pair excitation is zero when only one constituent of a particle-hole pair is excited (i.e., either $b_\mu$ or $b_\nu$ is zero), even when there are mixed particles and holes from different pairs in the system; it is non-zero when both constituents of a particle-hole pair are excited, even if there is only one such pair present in the system [see $(\Pi)$ in Fig.~\ref{fig:Hs}(d), for example]. A special case occurs when the system has a zero mode $\psi_0$ that is mapped by pseudo-chirality to its left eigenstate $\bar{\psi}_0$. We consider it as both a particle and a hole, and $(A)$ is also finite even when just this zero mode is excited.}

To exemplify this disparity and the pair excitation indicated by $(A)$, we consider the $n\times n$ topological photonic lattice shown in Fig.~\ref{fig:topo}(a). It has the same coupling $g$ in the vertical direction, and the asymmetric couplings in the horizontal direction are constants in each row, with their phases increased by $\pi/2$ successively in the vertical direction. This configuration can be realized using spatially displaced ring couplers \cite{Hafezi1,Hafezi2,topolaser1,topolaser2,toporouting}, and it leads to a synthetic gauge field for photons, with each of the smallest plaquettes pierced by a flux of $\pi/2$.

In the absence of on-site potentials, the underlying Hermitian system has sublattice symmetry, and its band structure is shown in Fig.~\ref{fig:topo}(b) with a Dirac cone at the $\Gamma$ point. \cc{With an \textit{arbitrarily} gain and loss landscape imposed, sublattice symmetry is lifted and it was unclear whether this type of non-Hermitian topological insulators still has symmetry protection \cite{Kawabata}, either in the form of $\e_\mu=-\e_\nu$ or $\re{\e_\mu}=-\re{\e_\nu}$. No non-Hermitian chiral symmetries have been reported in this case, and the NHPH symmetry identified in Refs. \cite{zeromodeLaser,Defect}, which exists in a wide range of non-Hermitian systems and leads to $\re{\e_\mu}=-\re{\e_\nu}$, does not appear here due to the asymmetric (albeit Hermitian) couplings.}

\cc{Nevertheless, this NHPH symmetry can be viewed as a special form of pseudo-anti-Hermiticity \cite{Scolarici}, i.e.,
\be
\eta H^\dagger \eta^{-1} = -H, \label{eq:anti-pseudoH}
\ee
which takes the form $\{\eta K,H\}=0$ ($K$: complex conjugation) when $H$ is symmetric. Remarkably, we find that this non-Hermitian topological lattice in fact possesses pseudo-anti-Hermiticity, even when the gain is just along the edges or random (as in the recent experiments of topological insulator lasers \cite{topolaser1,topolaser2,toporouting}). Pseudo-anti-Hermiticity leads to $\e_\mu=-\e_\nu^*$ [Fig.~\ref{fig:topo}(d)], or equivalently, $\re{\e_\mu}=-\re{\e_\nu}$, and its symmetry operator is given by $\eta=P_A-P_B\equiv C$ \cite{SM}, where $P_{A,B}$ are the projection operators onto the two sublattices indicated by filled and open dots in Fig.~\ref{fig:topo}(a). It is easy to show that pseudo-chirality is then present alongside $PT$ symmetry, with the former specified by $A = PC$.}

\begin{figure}[b]
\includegraphics[clip,width=\linewidth]{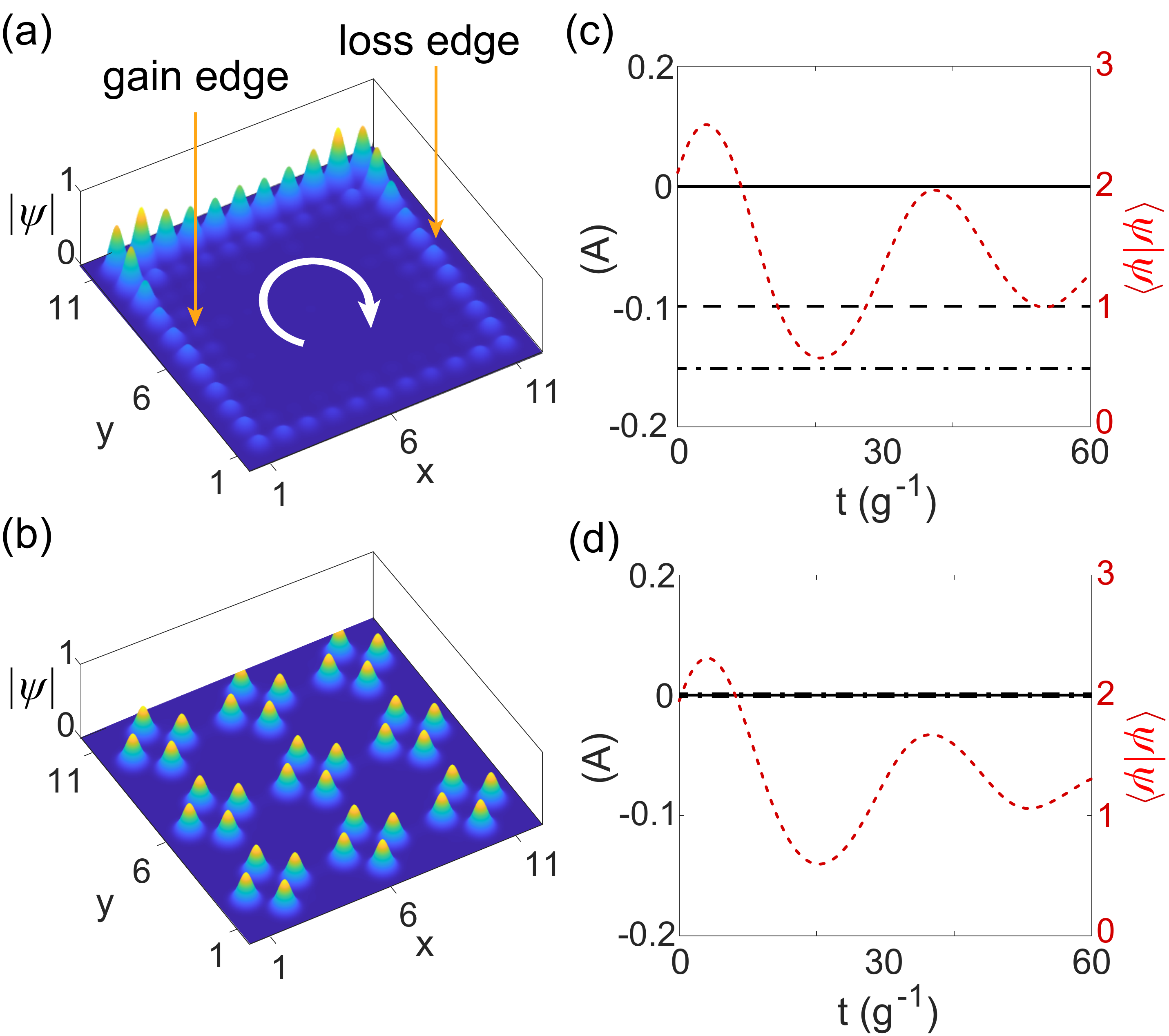}
\caption{Constant of motion in a topological lattice with the Landau gauge. (a,b) A CW chiral edge mode and the only zero mode ($\omega_\mu=0$) in the system with $n=11$, respectively. (c) Constant of motion $(A)$ for a pure CW chiral edge state (solid). Dashed and dash-dotted lines show, respectively, its values when the zero mode and the CCW chiral edge state is also excited. (d) Same as (c) but for $n=10$ where a zero mode does not exist.
} \label{fig:topo2}
\end{figure}

We realize this scheme by introducing a gain edge and a loss edge with on-site potential $\pm i\gamma$ on the left and right sides [Fig.~\ref{fig:topo}(a)], and the resulting complex spectrum, at $\gamma=g/10$, reflects the three aforementioned symmetries of this system [Fig.~\ref{fig:topo}(c)]. Below we focus on the bottom bandgap [BD1 in Fig.~\ref{fig:topo}(c)], where one chiral edge band emerges in a finite system \cite{topolaser1} just like in the Hermitian case. However, unlike the Hermitian case, the direction of a chiral edge state can now be easily identified by visualizing its spatial profile [Fig.~\ref{fig:topo2}(a)]: it is clearly a clockwise (CW) mode, because its intensity increases from bottom to top along the left (gain) edge and reduces from top to bottom along the right (loss) edge. % This is consistent with its direction calculated using a ribbon, and this mode has negative energy. The opposite chiral edge mode has the maximal intensity along the lower edge and have the opposite (positive) energy.

Noether's theorem manifested by Eq.~(\ref{eq:Ehrenfest2}) indicates that $(A)$ is a constant of motion. \cc{Due to the structure of $A=PC$, we find that $A$ is anti-symmetric (symmetric) when this lattice size $n$ is even (odd) \cite{SM}. As a result, $(A)$ vanishes universally when $n$ is even but indicates the pair excitation of particles and holes when $n$ is odd, even though their spectra are essentially the same with the $n$-odd case having a zero mode [Fig.~\ref{fig:topo2}(b)].}

\cc{A special observation based on these predictions is that $(A)$ now differentiates the excitation of a pure CCW or CW chiral edge propagation (``state'') versus its co-propagating counterpart. This finding is illustrated in Fig.~\ref{fig:topo2}(c) for $n=11$, where the solid line shows that $(A)=0$ is a constant of motion for the evolution of a CW wave packet (see the movie in Ref.~\cite{SM}). We note that although the wave packet consists of CW chiral edge modes that all have a \textit{real} frequency, its power (dotted line) still oscillates with time as a consequence of non-Hermiciticy \cite{NPreview}. If we also include the zero mode shown in Fig.~\ref{fig:topo2}(b) in the temporal evolution, now $(A)$ becomes nonzero [dashed line in Fig.~\ref{fig:topo2}(b)] since it is both a particle and a hole as mentioned before. Besides the zero mode, $(A)$ also become finite when we excite both chiral edge bands (dash-dotted line), which consists of pairing particles and holes. The same observations hold in the presence of symmetry-preserving disorder, as we show in Ref.~\cite{SM} by removing the lattice site at the center of the top row.
The universal vanishment of $(A)$ for the $n=10$ case is shown in Fig.~\ref{fig:topo2}(d).}

In summary, we have revealed and analyzed previously overlooked constants of motion in non-Hermitian systems, using the extension of Noether's theorem. Our systematic study of the latter has led to the introduction of a generalized symmetry, i.e., pseudo-chirality, which replaces the standard symmetries in quantum mechanics defined by a commutation relation. These results apply not only to non-Hermitian systems with gain and loss but also those with asymmetric hoppings \cite{SM}.  %We note that a nontrivial constant of motion due to pseudo-chirality exists in the simple case where the Hamiltonian is antisymmetric. $A$ in Eq.~(\ref{eq:pseudoC}) is then just the identity matrix. However, it then follows that all the diagonal elements of this Hamiltonian needs to be zero, which discards a powerful tool in non-Hermitian photonics and related fields, i.e., gain and loss in the form of imaginary on-site potentials \cite{NPreview,NPhyreview} which we have utilized in Figs.~\ref{fig:PT} and \ref{fig:prod}.
Our observation provides a suitable platform to investigate underlying symmetries from the dynamics of non-Hermitian systems, independent of whether such symmetries are spontaneously broken. It may also find applications in identifying and improving certain characteristics of topological systems, memory storage and information processing, where dissipation eliminates constants of motion found in Hermitian theories.

This project is supported by the NSF under Grant No. PHY-1847240.

%Different from quantum mechanics, now an eigenstate of the Hamiltonian is no longer a stationary state if these symmetries are spontaneously broken. (Check: if the system does not have pseudo-chirality but is chiral, will an eigenstate of H be a stationary state.)
%We also note that when the expectation value is taken in an eigenstate $\psi_\mu$ of $H$, then
%\be
%\frac{\partial \langle A\rangle}{\partial t} = \frac{\e_\mu-\e_\mu^*}{i\hbar}\langle A\rangle \label{eq:stationary1}
%\ee
%indicates that $\psi_\mu$ is a stationary state (i.e., for any linear operator $A$) only when $\e_\mu=\e_\mu^*$, requiring that pseudo-Hermiticity is not spontaneously broken in $\psi_\mu$ \cite{pseudoH_arxiv}.
%Similar to Eq.~(\ref{eq:stationary1}), we also point out that $\psi_\mu$ is a stationary state with respect to the inner product (\ref{eq:inner2}) and for any linear operator only when pseudo-chirality is not spontaneous broken, requiring $\e_\mu=-\e_\mu=0$ and hence $\psi_\mu$ to be a zero mode:
%\be
%\frac{\partial (A)}{\partial t} = \frac{2\e_\mu}{i\hbar}(A). \label{eq:stationary2}
%\ee

\end{document}